\documentclass[aps,prd,twocolumn,superscriptaddress,nofootinbib,amsmath,amssymb,preprintnumbers,floatfix,10pt]{revtex4-2}
\usepackage{graphicx}
\usepackage{amsmath,amssymb}
\usepackage{hyperref}
\hypersetup{
    colorlinks=true,
    citecolor=blue,
    linkcolor=magenta,
    filecolor=magenta,
    urlcolor=blue,
}
\usepackage{float}
\usepackage{slashed}
\usepackage{tikz}
\usepackage{tikz-feynman}
\usepackage{feynmp}
\usepackage{subfig}
\usepackage{gensymb}
\usepackage{multirow}
\usepackage[toc,page]{appendix}
\usepackage{aas_macros}
\usepackage{physics}
\usepackage{url}
\tikzfeynmanset{compat=1.1.0}

\newcommand{\tm}{\textrm}
\newcommand{\be}{\begin{equation}}
\newcommand{\ee}{\end{equation}}
\newcommand{\bs}{\begin{split}}
\newcommand{\es}{\end{split}}
\newcommand{\LL}{\mathcal{L}}

\newcommand{\OO}{\mathcal{O}}
\newcommand{\mx}{m_{\chi}}

\newcommand{\mev}{\textrm{ MeV}}
\newcommand{\gev}{\textrm{ GeV}}

\newcommand{\Nx}{N_{\chi}}

\begin{document}

\title{Stringent Constraints on Gravitational Wave Signatures of Dark Electromagnetism in Neutron Star Binaries}

\author{Ian Harris}
\email{ianwh2@illinois.edu}
\affiliation{Department of Physics, University of Illinois Urbana-Champaign, Urbana, Illinois 61801, U.S.A.}
\affiliation{Illinois Center for Advanced Studies of the Universe, University of Illinois Urbana-Champaign, Urbana, Illinois 61801, U.S.A.}

\author{Yonatan Kahn}
\email{yf.kahn@utoronto.ca}
\affiliation{Department of Physics, University of Toronto, Toronto, ON M5S 1A7, Canada}
\date{\today}

\begin{abstract}
    Gravitational wave interferometers have studied compact object mergers and solidified our understanding of strong gravity. Their increasing precision raises the possibility of detecting new physics, especially in a neutron star binary system that may contain hidden-sector particles. In particular, a new vector force between binary constituents, giving rise to dark electromagnetic phenomena, could measurably alter the inspiral waveforms and thus be constrained by gravitational wave observations. In this work, we critically examine the mechanisms for neutron stars to acquire enough hidden-sector particles with requisite couplings to furnish a detectable signature from dark electromagnetism. We demonstrate that the repulsive nature of vector forces imposes stringent constraints on any putative particle physics model or astrophysical environment which could give rise to such gravitational signatures. We argue that absent an extreme fine-tuning of parameters, such signatures are well out of reach of any current or near-future gravitational wave observatory.
\end{abstract}

\maketitle

\section{Introduction}
Dark matter (DM) has long eluded detection. For a single species feebly interacting with the Standard Model (SM), terrestrial experiments are rapidly closing in on their sensitivity limits, particularly for the MeV to TeV mass range~\cite{Akerib:2022ort,Essig:2022dfa}. However, astrophysical observations can provide complementary probes \cite{Drlica-Wagner_2022}. Compact objects, in particular neutron stars (NS), are an attractive prospect for DM studies \cite{battaglieri2017cosmicvisionsnewideas, cirelli2024darkmatter}, given their large density and strong gravitational fields that can accrete and/or focus large quantities of ambient DM \cite{Goldman_1989, Bell_2020, Bell_2021, DeRocco_2022, Busoni_2022, Kouvaris_2010, de_Lavallaz_2010, Nelson_2019, Leane_2021}. DM can affect NS properties such as tidal deformability \cite{Arvikar_2025, Rafiei_Karkevandi_2023, Ellis_2018, CIANCARELLA2021100796}, radial \cite{gleason_2022} and $f$-mode \cite{flores_2024, das_2021, shirke_2024} oscillations, as well as thermal signatures \cite{Acevedo_2020, Garani_2021, avila_2024, Bell_2024}, all of which have been used to place constraints on particular models.

Long range DM-DM self-interactions (specifically, a Yukawa-type force with a range at least as large as typical NS-NS binary separations) can potentially lead to observable signatures in NS-NS binaries, with a character that depends on the spin of the mediating particle. Long-range spin-2 forces are effectively modifications to general relativity and are difficult to incorporate in a particle physics framework. Scalar (spin-0) forces are universally attractive, and have been studied in Refs.~\cite{colpi_1986, Alexander_2018, rutherford_2023, PhysRevD.105.023001, cong2024spindependentexoticinteractions, jockel_2023}. Vector (spin-1) interactions, arising from U(1) gauge forces, uniquely lead to both attractive and repulsive interactions. A hidden sector containing such a U(1) force can thus furnish NS-NS binaries with dark analogues of the full spectrum of electromagnetic (EM) phenomenon, including electric and magnetic forces and electric and magnetic dipole radiation. These can potentially give unique imprints in the gravitational wave (GW) signature from the merger, as electric and magnetic phenomena depend differently on the relative velocity of the NSs in the binary. As shown in Ref.~\cite{owen_2025}, the strength of the GW signal of such dark EM scales (in particle physics units with $\hbar = c = k_B = 1$) as
\be
S\equiv \frac{\Nx^2g_D^2}{G_NM_*^2},
\label{eq:signal}
\ee
where $\Nx$ is the number of DM particles in a NS, $g_D$ is the dark U(1)$_D$ gauge coupling, $G_N$ is Newton's constant, and $M_*$ is the NS mass. Ref.~\cite{owen_2025} demonstrated that the current design of LIGO is sensitive to dark EM at a threshold strength $S \gtrsim 10^{-3}$; importantly, the same signal strength parameter governs both the effects of dark EM forces and dark dipole radiation.

In this work, we systematically study the conditions required to achieve observable signals of dark vector forces at current or future gravitational wave observatories. We find that the only such scenarios which survive laboratory and astrophysical constraints are so contrived or finely-tuned as to push the boundaries of plausibility. The principal reason for these stringent constraints was first noted in Ref.~\cite{Kopp_2018}, published shortly after the first observation of an NS-NS merger GW170817~\cite{Abbott_2017}: a long-range NS-NS force requires each NS to have a net dark charge, leading to a strong repulsive dark electrostatic force which competes with gravity. We extend the arguments of Ref.~\cite{Kopp_2018} to include models which give smaller signal strengths than $S \sim 10^{-2}$ (which was the sensitivity at the time of GW170817), and find that these models are still effectively ruled out. While it is important to keep an open mind as to what new physics signals may be lurking in GW observations, the goal of our work is to demonstrate the severely restrictive nature of the hypothesis of a spin-1 force, which also has implications for complementary astrophysical and laboratory searches for dark forces.

This paper is organized as follows. In Section \ref{sec:gravstab}, we review constraints on dark gauge couplings from gravitational stability and their implications for LIGO sensitivity. In Section \ref{sec:accretion}, we discuss limitations on ambient dark matter particles captured by gravitational infall and scattering. Section \ref{sec:thermalproduction} discusses thermal production of dark-sector particles in the progenitor supernova and derives a maximum GW signal strength from stability constraints in the presence of the new force responsible for thermal production. In Section \ref{sec:neutrondecay}, we examine the hypothetical phenomenon of neutron ``dark'' decay and show that existing models of this scenario which manage to evade strong laboratory and astrophysical constraints are nonetheless inconsistent with a new dark vector force. We conclude in Section \ref{sec:conclusion}.

\section{Gravitational stability constraints}
\label{sec:gravstab}

In order to have a long-range dark force between NSs in a binary, or to source dark EM radiation, at least one NS in the binary must have a net dark charge, $Q_D = N_\chi g_D$.\footnote{A possible loophole is if the NS has a net-neutral dark charge, comprised of equal numbers of species of opposite charges. In that case, any dark charge separation in the NS (which would of course require its own dynamical explanation) would lead to $1/r^4$ dipole-dipole forces in the binary, which are weaker than the $1/r^2$ forces studied in Ref.~\cite{owen_2025} but can be constrained within the post-Newtonian framework for modifications to general relativity~\cite{Yunes:2009ke,Cornish:2011ys,Yunes:2016jcc}. To our knowledge we are not aware of any previous work studying dipole-dipole dark sector forces in NS binaries. Nevertheless, we leave this option open for future study.} Without loss of generality we have taken the DM particles to have charge 1 in units of the U(1)$_D$ gauge coupling $g_D$. Assuming this U(1)$_D$ is the only long-range force in the dark sector, the NS then sources a dark electrostatic potential, whose value at the radius of the NS $R_*$ is
\begin{equation}
    \Phi_D = \frac{Q_D}{R_*}.
\end{equation}
The gravitational potential of the NS at the surface is
\begin{equation}
    \Phi_G = -\frac{G_N M_*}{R_*}.
\end{equation}
For the charge configuration $Q_D$ to be stable, the gravitational potential energy $|U_G| = m_\chi |\Phi_G|$ for a DM particle of mass $m_\chi$ must exceed the electrostatic potential energy $|U_D| = g_D |\Phi_D|$, otherwise the electrostatic repulsion will overwhelm the gravitational attraction and DM particles will be repelled away from the star. Thus, we must have
\begin{equation}
\label{eq:gravstab0}
    g_D|\Phi_D| \leq m_\chi |\Phi_G| \implies N_\chi g_D^2 \leq G_N M_* m_\chi.
\end{equation}
Multiplying both sides by $\frac{N_\chi}{G_N M_*^2}$ gives the LIGO signal strength $S$ on the left-hand side, yielding
\begin{equation}
\label{eq:gravstab}
    S \leq \frac{N_\chi m_\chi}{M_*} \equiv f_\chi,
\end{equation}
where the right-hand side $f_\chi$ is the mass fraction of DM inside the NS. This relation (essentially identical to that first derived in Ref.~\cite{Kopp_2018}, but here expressed in a model-independent way) implies that \emph{the maximum achievable LIGO signal strength for a single long-range spin-1 DM force is bounded by the mass fraction of DM in a NS.} To achieve $S \sim 10^{-3}$, at least one NS must contain of order $0.1\%$ DM. In the following sections, we will see that this condition is generically extremely difficult to realize in concrete models, even for $S$ well below the current LIGO sensitivity.

So far our arguments have been completely model-agnostic and independent of the process that results in the accumulation of DM inside the NS. One can of course consider additional model-building ingredients as loopholes to the gravitational stability argument; for example, adding an additional attractive long-range DM-DM force from exchange of spin-0 mediators. However, for two NSs with like U(1) charges, such an additional scalar force would only serve to reduce the observable GW signal, since the same attractive force acting between NSs would partially cancel the repulsive U(1) force. A scenario where the two NSs in a binary have large and opposite U(1)$_D$ charges seems extremely implausible, requiring exquisitely fine-tuned initial conditions to preferentially accumulate one DM species in one NS and another oppositely-charged species in the other. While we are not definitively ruling out such model-building contortions, our goal here is to illustrate the lengths to which one must go to evade the very generic constraint in Eq.~(\ref{eq:gravstab}).

\section{Accretion}
\label{sec:accretion}
Strong gravitational fields from compact objects can focus ambient galactic DM, and the large SM density in the compact object can provide a large number of targets for DM-SM scattering, resulting in DM capture. However, we will show that even the enormous density of SM matter in a NS is not sufficient to accrete a sufficient mass of DM to produce an observable GW signal, absent a comparably enormous local overdensity of DM in the neighborhood of the NS.

\subsection{Geometric capture}

Regardless of the particle physics model of DM, the maximal capture rate occurs when the DM-SM cross section saturates the geometric cross section of the NS~\cite{Bell:2018pkk,Bell_2020}
\begin{equation}
\label{eq:Cgeom}
    C_{\rm geom.} = \frac{\pi R_*^2(1 - B(R_*))}{v_* B(R_*)}\frac{\rho_\chi}{m_\chi}{\rm erf}\left(\sqrt{\frac{3}{2}}\frac{v_*}{v_d}\right),
\end{equation}
where $B(R_*) = 1 - \frac{2G_NM_*}{R_*}$, $\rho_\chi$ is the local density of DM far from the NS, $v_*$ is the peculiar velocity of the NS, and we have assumed DM has a Maxwellian velocity distribution with dispersion $v_d$. Note that Eq.~(\ref{eq:Cgeom}) already accounts for all gravitational focusing and general relativistic effects.

For a constant capture rate given by geometric capture, the number of captured DM particles scales linearly with the lifetime of the binary $t_{\rm lifetime}$ and inversely with the DM mass $m_\chi$. Taking reasonable NS parameters $R_* = 10 \ {\rm km}$ and $M_* = 2 M_\odot$, which only affect the geometric cross section by $\mathcal{O}(1)$ factors, gives\footnote{We do not explicitly show dependence on the erf factor (which results from integrating over the Maxwellian distribution), since it also only contributes an $\mathcal{O}(1)$ factor.}
\be
\begin{split}
\Nx\simeq & \ 2\times10^{43}\left(\frac{1\gev}{\mx}\right)\left(\frac{220 \ {\rm km/s}}{v_*}\right)\left(\frac{t_{\tm{lifetime}}}{10\tm{ Gy}}\right)\\  &\times \left(\frac{\rho_{\chi}}{0.4 \ {\rm GeV/cm}^3}\right).
\end{split}
\ee
The DM mass fraction then scales as
\be
f_\chi \simeq  7 \times10^{-14}\left(\frac{220 \ {\rm km/s}}{v_*}\right)\left(\frac{t_{\tm{lifetime}}}{10\tm{ Gy}}\right)  \left(\frac{\rho_{\chi}}{0.4 \ {\rm GeV/cm}^3}\right),
\ee
which is independent of the DM mass and depends only on the local DM density $\rho_\chi$.

Regardless of the particle physics model, accretion via geometric capture can only accumulate a dark mass fraction of $\OO(10^{-13})$ in the age of the universe, assuming the local density of DM in the neighborhood of the NS is comparable to its value in the solar neighborhood. This is clearly insufficient to yield an observable gravitational wave signal -- which by our arguments in Sec.~\ref{sec:gravstab} is bounded by $S \lesssim 10^{-13}$ -- at any current or near-future experiment. In fact, DM with long-range self-interactions, whether a single species \cite{McCullough_2013, Alonso_alvarez_2024, Agrawal_2017, Kaplinghat_2016, graham2025cosmologicallimitsstrongdark, giffin2025structureformationdarkmagnetohydrodynamics, Cruz_2023, DeRocco_2025} or a multi-component dark sector such as atomic dark matter \cite{Kaplan:2009de,FAN_2013,Braaten_2018, Tulin_2018,Beauchesne_2021,Gurian_2022,roy_2024} is generically constrained to be an $\OO(5\%)$ subcomponent of cosmological DM. The density of accreting DM is thus strictly smaller than the total local density, further restricting the viable parameter space.

\subsection{Accretion loopholes}

There are two possibilities for increasing $f_\chi$: self-capture and extremely large local overdensities of DM. The former still fails to achieve detectable signals, while the latter requires astrophysical conditions which strain plausibility.

\subsubsection{DM Self-Accretion Timescale}
Since by assumption DM is charged under a long-range U(1) force, the same force leads to Coulomb-like DM self-interactions. DM may thus be ``self-captured'' by accumulating off of an initial ``seed" of $N_0$ DM particles in the NS. The self-capture rate is proportional to the DM population in the star, so the self-accretion is exponential,
\begin{equation}
    N_{\chi}(t)=N_0\exp\left[\sigma_{\chi\chi}\frac{(1-B(R_*))}{v_{*}B(R_*)}{\rm erf}\left(\sqrt{\frac{3}{2}}\frac{v_*}{v_d} \right)n_{\chi} t\right],
    \label{eq:self-acc}
\end{equation}
where $n_{\chi} = \rho_\chi/m_\chi$ is the local DM number density. The capture rate accounts for gravitational focusing, consistent with Eq.~(\ref{eq:Cgeom}). The differential cross section for DM-DM scattering is the M\o ller cross section in the non-relativistic limit:
\be\begin{split}
\frac{d\sigma_{\chi\chi}}{d\Omega}=&\frac{g_D^4}{64\pi^2p^4\left(p^2+\mx^2\right)\sin^4\theta} \bigg[4\left(\mx^2+2p^2\right)^2+ \\
&\left(4p^4-3\left(\mx^2+2p^2\right)^2\right)\sin^2\theta+p^4\sin^4\theta \bigg],
\end{split}
\ee
with $p$ and $\theta$ the center-of-mass momentum and scattering angle. To be captured, the incoming $\chi$ must lose its virial kinetic energy, given by $\simeq\frac{1}{2}\mx v_d^2$. This sets a minimum value for $\theta$, regulating the collinear singularity that arises in Coulomb scattering processes. Integrating the cross section over scattering angles which lead to capture yields
\be
\begin{split}
\sigma_{\chi\chi}\simeq& \ 3\times10^{-23}\tm{cm}^2\left(g_D\right)^4\left(\frac{\gev}{\mx}\right)^2.
\end{split}
\label{eq:identicalscatter}
\ee
On the other hand, squaring the gravitational stability constraint Eq.~(\ref{eq:gravstab0}) and rearranging gives
\be
\frac{g_D}{m_\chi} \leq \frac{G_N}{\sqrt{S}},
\label{eq:ratiolim}
\ee
which bounds the DM charge-to-mass ratio in terms of the potential signal strength $S$ today once the DM cloud has fully accumulated.
We can therefore rewrite (\ref{eq:identicalscatter}) as 
\be
\begin{split}
    \sigma_{\chi\chi}\lesssim 10^{-99}\ {\rm cm}^2\left(\frac{\mx}{\rm{GeV}}\right)^2 \times \frac{1}{\sqrt{S}}.
\end{split}
\label{eq:sigmavsmassscaling}
\ee
The duration of an $e$-fold of DM accumulation therefore scales as
\be
\begin{split}
t_{e\tm{-fold}}
\gtrsim& \ 4.5\times10^{49} \ t_{H} \times \sqrt{S}\\
&\times\left(\frac{v_*}{100\tm{ km/s}}\right)\left(\frac{10^{18}\gev/\tm{cm}^3}{\rho_{\chi}}\right)\left(\frac{\rm{GeV}}{\mx}\right).
\label{eq:selfscatterscale}
\end{split}
\ee
Even taking the ambient DM density to an extremely large value (as we discuss further below) cannot bring this timescale below the age of the universe $t_H$, unless the eventual signal strength $S$ is unmeasurably small.

\subsubsection{Galactic Central Overdensity}
One option remains: postulating an extraordinary DM overdensity in the local neighborhood of the NS. If such conditions were to occur, they would very likely be at the center of a galaxy. Globular cluster modeling gives an ambient DM density increase of four orders of magnitude at the galactic center \cite{posti_2018}. Stellar densities in elliptical and spiral galaxies have been observed to scale as steeply as $\rho_{\tm{stellar}}\sim r^{-2.5}$ \cite{Merritt_2004}; for $\rho_{\tm{DM}}\sim r^{-2.33}$, the DM density at the galactic center would reach $10^{12}\gev/\tm{cm}^3$ \cite{Bertone_2005}. More recent, fully relativistic treatments postulate either a $\sim2\times10^{19}\gev/\tm{cm}^3$ DM ``spike"  \cite{Sadeghian_2013} or a $\sim2\times10^{18}\gev/\tm{cm}^3$ ``mound" \cite{Bertone_2025}; such a distribution builds up adiabatically to these densities and allows DM to cluster closer to the Schwarzschild radius of the central supermassive BH. Indeed, the presence of rapidly decaying black hole binaries motivates such spikes \cite{scarcella2025hintsdarkmatterspikes}. A central spike of $10^{19}\gev/\tm{cm}^3$ would be sufficient to obtain $f_\chi \sim \OO(1)$ in a NS binary lifetime of 1 Gyr, and thus a detectable gravitational wave signature.

These enormous overdensities, however, could be destroyed as a result of several effects. The first is scouring, which has long been postulated as a mechanism in which a supermassive black hole binary system attracts the DM out of the spike \cite{Milosavljevic_2002,Ravindranath_2002}. Another is gravitational wave recoil, proposed more recently to explain the formation of larger galactic cores \cite{khonji2024coreformationbinaryscouring, 10.1093/mnras/stab435}. Furthermore, gravitational scattering between DM and stars has been predicted to deplete the spike over a period of 10 Gy, even if no BH binary is present from a prior galactic merger \cite{Ilyin_2004, Gnedin_2004, Merritt_2004}. A candidate binary system, even one that achieves geometric capture, must be located at the center of a galaxy with a large DM overdensity in the absence of these depleting effects.

\section{Thermal Production}
\label{sec:thermalproduction}

If DM is not accreted in the NS, it must be produced in sufficient quantities during the formation of the NS. In this section, we examine thermal production in the progenitor supernova as another possibility to create enough DM to yield a detectable GW signature. This production channel has previously been proposed in Refs. \cite{Kopp_2018, Nelson_2019}. Here we show that the addition of a long-range vector force, combined with a modified version of the gravitational stability considerations of Sec.~\ref{sec:gravstab}, also render this model unviable.

\subsection{Model construction}
The additional necessary ingredient for this model is an interaction between baryons and DM to produce the DM in a supernova. Assuming U(1)$_D$ charge conservation, and that the NS progenitor is uncharged under U(1)$_D$, the DM must be pair produced from baryons $N$ via $N \to N \bar{\chi} \chi$. The resulting NS would then be neutral under U(1)$_D$ and lead to no GW signatures unless there were a mechanism to preferentially attract $\chi$ and repel $\bar{\chi}$ (or vice versa). Ref.~\cite{Nelson_2019} noted that the same interaction which produces $\bar{\chi} \chi$ could provide the required asymmetric force if $\chi$ were attracted to baryons, in which $\bar{\chi}$ would be repelled from baryons and thus expelled from the NS.

\begin{figure}[t]
    \includegraphics[width=0.95\linewidth]{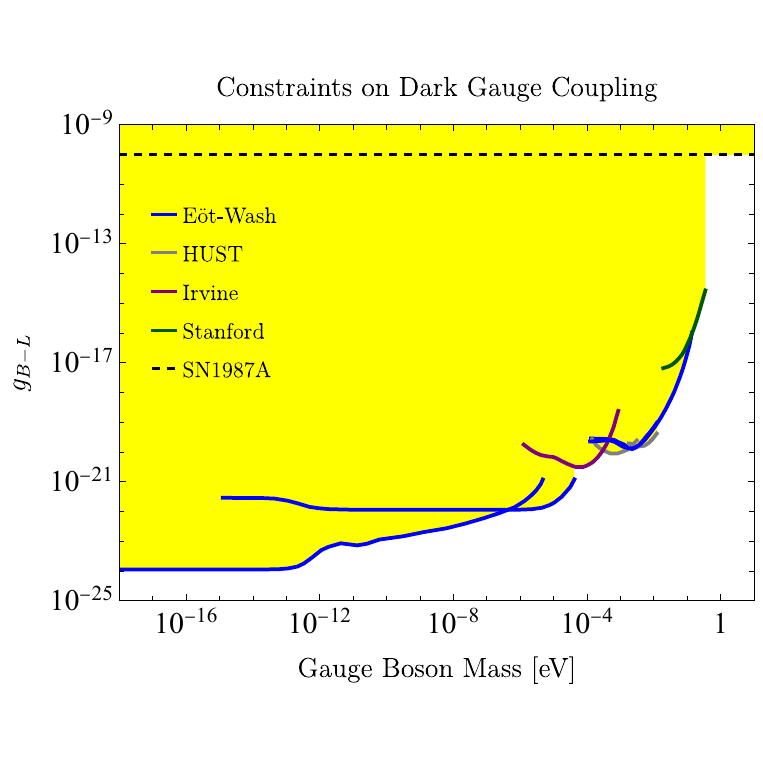}
    \vspace{-0.5cm}
    \caption{Experimental constraints taken from \cite{Wagner_2012, Berg__2022, Kapner_2007} and adjusted to show the excluded region in terms of the gauge coupling and boson mass. Solid lines show fifth force constraints; the dashed line shows the limit from supernova bremsstrahlung production.}
    \label{fig:fifthforce}
\end{figure}

We are thus led to a model which permits the process $N \to N A' \to N \bar{\chi} \chi$, and where the NS also sources a large classical scalar potential $\Phi$ to attract $\chi$ and repel $\bar{\chi}$. In order to get both attractive and repulsive interactions, $A'$ must be a vector force. One possibility is that this is the same $A'$ giving the long-range interaction which results in our desired GW signature. To avoid stringent constraints on the production of longitudinal modes of light vector bosons \cite{Dror:2017ehi,Dror:2017nsg}, the DM-baryon force should be anomaly-free: the only option is baryon minus lepton number ($B-L$). Long-range $B-L$ forces, which effectively couple to neutron number, have been extensively studied in fifth-force experiments. Fig.~\ref{fig:fifthforce} shows a summary of these constraints, adapted from Refs.~\cite{Wagner_2012, Berg__2022, Kapner_2007}. Requiring that the dark EM force have a $\gtrsim {\rm km}$ range implies that $g_{B-L} \lesssim 10^{-24}$; such a small gauge coupling is not sufficient to create the required energy difference to preferentially attract $\chi$ over $\bar{\chi}$~\cite{Nelson_2019}.

One can accommodate a weak and/or short-range DM-baryon force by kinetically mixing it with the long-ranged U(1)$_D$ dark-sector force. For concreteness, we take the U(1)$_D$ force to be massless and consider the Lagrangian following diagonalization of the gauge kinetic terms:
\begin{widetext}
    \be
\begin{split}
    \LL\supset & \bar{\chi}\left(i\slashed{\partial}-g_D\left(\slashed{A}_{D}+\frac{\epsilon}{\sqrt{1-\epsilon^2}}\slashed{A}_{B-L} \right)-\mx\right)\chi + g_{B-L}A^{\mu}_{B-L}J^{B-L}_{\mu}\\
    &-\frac{1}{2}m^2_{B-L}A^{\mu}_{B-L}A_{\mu}^{B-L}-\frac{1}{4}F^{\mu\nu}_{B-L}F_{\mu\nu}^{B-L}-\frac{1}{4}F_D^{\mu\nu}F^D_{\mu\nu},
\end{split}
\label{eq:adjLProgenitor}
\ee
\end{widetext}
where $\epsilon < 1$ is the kinetic mixing parameter, $m_{B-L}$ is the $B-L$ gauge boson mass, and $J_\mu^{B-L}$ is the $B-L$ current. 

\subsection{GW signal constraints}

Ref.~\cite{Kopp_2018} argues that progenitor supernovae can produce up to $5.5\times10^{-4}M_{\odot}$ of $\bar{\chi}\chi$ pairs, provided that their mass does not considerably exceed the progenitor supernova temperature of $\sim50\mev$ \cite{Nelson_2019}. Despite the large value of $f_\chi$, we show that this model is still not viable because of existing constraints on $B-L$ forces.

The baryons in the NS will contribute an attractive potential energy which may partially counteract the self-repulsion from the DM cloud, and which may be calculated from mean-field theory. Neglecting the kinetic term, the terms in the Lagrangian which depend on $A^{\mu}_{B-L}$ are
\be
    \LL \supset g_{B-L}A^{\mu}_{B-L}J^{B-L}_{\mu}-\frac{1}{2}m^2_{B-L}A^{\mu}_{B-L}A_{\mu}^{B-L}.
\ee
Applying the classical Euler-Lagrange equation for $A^{\mu}_{B-L}$ 
and solving for the gauge field gives
\be
A^{B-L}_{\mu}= \frac{g_{B-L}}{m^2_{B-L}} J^{B-L}_{\mu}.
\ee
We now take the non-relativistic mean-field limit, $A^{\mu}_{B-L}\xrightarrow[]{}(A^0_{B-L},\vec{0})$, which lets us write $A^0_{B-L}$ in terms of the baryon density $n_b$:
\be
A^0_{B-L}=\frac{g_{B-L}}{m^2_{B-L}} J_{B-L}^0 =\frac{g_{B-L}}{m^2_{B-L}}n_b.
\ee
This acts as a potential for the DM density $\bar{\chi} \gamma^0 \chi \equiv n_\chi$, where we assume that the U(1)$_D$ charge of $\chi$ is negative to give an attractive force after taking into account kinetic mixing.

The total potential energy of a test $\chi$ particle at radius $R_*$ is now a sum of three terms,
\begin{align}
U_{\rm tot} & = U_D + U_G + U_{B-L} \\
& = \frac{g_D^2 N_\chi}{R_*} - \frac{G_N M_* m_\chi}{R_*} - \frac{\epsilon g_D g_{B-L}}{\sqrt{1-\epsilon^2}}\frac{n_b}{m_{\rm eff}^2},
\end{align}
where
\be
m_{\rm eff} \simeq {\rm max}[m_{B-L},R_*^{-1}]
\ee
because for $m_{B-L} < 1/R_*$, the mean-field approximation breaks down but the potential energy simply becomes the Coulomb energy of a long-range force. Since we must have $m_\chi \lesssim 50 \ {\rm MeV}$, the maximum value of the gravitational potential energy for a NS of mass $M_* = 2 M_\odot$ and radius 10 km is $|U_G| \simeq 15 \ {\rm MeV}$. On the other hand, maximizing the attractive potential energy of the baryons by taking $\epsilon \sim 1/\sqrt{2}$, $g_D \sim 1$, $n_b \sim 1 \ {\rm fm}^{-1}$, $m_{\rm eff} \sim (12 \ {\rm km})^{-1}$ and $g_{B-L} \sim 10^{-10}$ gives $|U_{B-L}| \sim 2.8 \times 10^{30} \ {\rm MeV}$. This means that for a wide range of parameters, the stability of the $\chi$ cloud is completely dominated by a competition between the repulsive U(1)$_D$ force and the attractive U(1)$_{B-L}$ force, with gravitational attraction a negligible contribution. 
\begin{figure}[t]
    \includegraphics[width=0.95\linewidth]{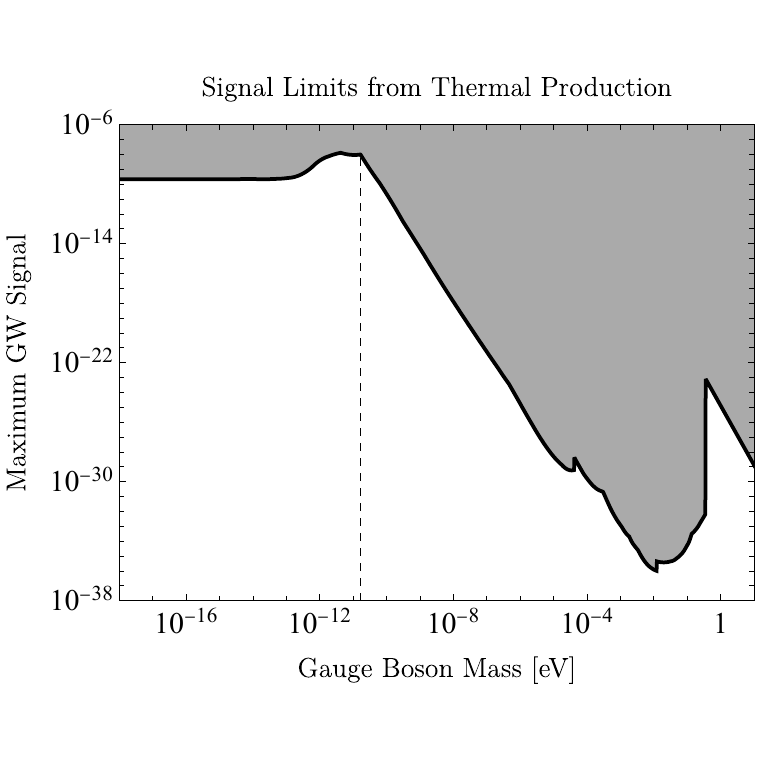}
    \vspace{-0.5cm}
    \caption{Maximum GW signal as a function of $B-L$ gauge boson mass $m_{B-L}$, for the thermal production scenario. The vertical dashed line marks where $m_{B-L}=R_*^{-1}$, where the mean-field calculation is no longer valid and the $B-L$ force is effectively long-ranged.}
    \label{fig:siglimstherm}
\end{figure}

Neglecting the gravitational attraction, the condition for stability is
\be
 \frac{g_D^2 N_\chi}{R_*} \leq \frac{\epsilon g_D g_{B-L}}{\sqrt{1-\epsilon^2}}\frac{n_b}{m_{\rm eff}^2}.
\ee
Substituting $g_D$ for the GW signal strength, $g_D = \frac{M_*}{N_\chi} \sqrt{S G_N}$, and setting $\epsilon = 1/\sqrt{2}$ as the maximum reasonable value of the kinetic mixing parameter (since larger values would imply a nearly-singular kinetic matrix), we find
\be
S \leq \frac{g_{B-L}^2n_b^2 R_*^2}{G_N M_*^2 m_{\rm eff}^4}.
\ee

Taking typical values for the NS parameters ($M_* = 1.4 M_\odot$, $R_* = 12 \ {\rm km}$, $n_b = 1 \ {\rm fm}^{-3}$) and setting $g_{B-L}$ to the maximum value for each $m_{\rm eff}$ allowed by Fig.~\ref{fig:fifthforce}, we obtain the upper bound on the GW signal over the $B-L$ parameter space shown in Fig.~\ref{fig:siglimstherm}. We have confirmed that for the entire boundary of the shaded region in Fig.~\ref{fig:siglimstherm}, $\left|U_{B-L}\right|$ is at least two orders of magnitude greater than $\left|U_G\right|$, so our calculation which neglects the gravitational potential energy is self-consistent. We see that over the entire range of possible $B-L$ gauge boson masses, the GW signal is no larger than $10^{-8}$, which is well outside the sensitivity of any current or near-future experiment.

\section{Neutron Decay}
\label{sec:neutrondecay}
The last production mode that we examine is via neutron ``dark'' decays~\cite{McKeen_2018,Cline_2018}, which was explored as a possible source of dark sector particles in NSs in~\cite{Kopp_2018}. Here we show that it is impossible to include a long-range vector force in models which have already been studied, which are already subject to extremely stringent constraints from, e.g., NS equations of state.

Our argument is as follows. By U(1)$_D$ charge conservation, either the initial neutron is charged under the new U(1)$_D$, or the final state of the decay is neutral under U(1)$_D$. The former scenario would imply that, even prior to any neutron dark decays, the entire NS is charged under a new long-range force. This possibility is ruled out by the long-range force constraints discussed in Sec.~\ref{sec:thermalproduction}. Therefore, the decay to the dark gauge boson, $N \to \chi A_D$, studied in Ref.~\cite{Cline_2018} for the case of heavy $A_D$, is not a viable production mechanism for dark charges when the U(1) dark force is long-ranged.

The alternative is for the decay to proceed as $N \to \chi \phi$, where the fermion $\chi$ and scalar $\phi$ are new dark-sector particles with opposite U(1)$_D$ charges. This model was studied in Ref.~\cite{Grinstein_2019}, with Lagrangian
\be
\begin{split}
\LL_{\tm{scalar decay}}\supset&\lambda_q\epsilon^{ijk}\bar{u}^c_{Li}d_{Rj}\Phi_k+\lambda_{\chi}\Phi^{*i}\bar{\tilde{\chi}}d_{Ri}\\
&+\lambda_{\phi}\bar{\tilde{\chi}}\chi\phi+\mu H^{\dagger}H\phi+g_{\chi}\bar{\chi}\chi\phi+\tm{h.c.}
\label{eq:scalardecay}
\end{split}
\ee
where $\Phi$ is a heavy color triplet scalar of hypercharge $-1/3$ and baryon number $-2/3$, and $\tilde{\chi}$ is another hidden fermion. The mass constraints already require a significant amount of fine-tuning~\cite{Grinstein_2019,basterogil_2024}:
\be
\begin{split}
937.993\mev&<\mx+m_{\phi}<939.563\mev,\\
937.993\mev&<m_{\tilde{\chi}},\\
|\mx-m_{\phi}|&<938.783\mev.
\end{split}
\label{eq:massconstraints}
\ee
However, the presence of the trilinear term $\mu H^\dagger H \phi$ and the Yukawa term $g_\chi \bar{\chi} \chi \phi$ imply that $\phi$ must be neutral under any new U(1), and thus this model is not viable with the addition of U(1)$_D$. 

Of course, we do not claim that this simple argument rules out \emph{all} possible models of dark neutron decay giving observable GW signatures. However, modifying low-energy physics while remaining consistent with observed NS phenomenology is extremely nontrivial, and the addition of a new long-range force may simply overconstrain the space of viable models.

\section{Conclusion}
\label{sec:conclusion}
In this paper, we have demonstrated the stringent constraints and highly contrived model-building required to engineer a detectable GW signature $S$ from long-range dark vector forces in a NS binary system. If DM accumulates through capture of the ambient galactic DM, the signal strength is bounded by $S \lesssim 10^{-14}$ for DM densities comparable to the local density. A NS can only accrete an observable amount of DM very close to the centers of the steepest galactic DM density profiles. While an NS could theoretically acquire this large dark mass fraction from thermal production in the progenitor supernova or neutron decay, a plethora of additional constraints preclude a detectable signal. In the thermal production scenario, fifth-force constraints restrict $S \lesssim 10^{-8}$. The neutron decay scenario already requires an extreme fine-tuning of the dark-sector particle mass to just below the neutron mass; we have shown that the models which have been studied thus far are impossible to reconcile with the addition of a long-range U(1)$_D$ force, and are likely susceptible to stringent laboratory tests in the near future.

The current sensitivity of LIGO to signatures of dark EM is $S \gtrsim 10^{-3}$~\cite{owen_2025}, at least 5 orders of magnitude weaker than that required to detect any of the production scenarios we have examined. While we do not fully exclude the possibility of finely-tuned models which give rise to larger signal strengths $S$, if a deviation from GR predictions is observed in NS inspirals localized anywhere but the center of a host galaxy, we argue that it is extremely unlikely to arise from dark vector forces.

\section*{Acknowledgments}
We thank Caroline Owen, David Curtin, Nicol\'{a}s Yunes, Jessie Shelton, Gil Holder, Keegan Humphrey, and Mateus Reinke Pelicer for helpful conversations. The work of I.H. and Y.K. was supported in part by DOE grant DE-SC0015655.  Y.K. acknowledges the support of a Discovery Grant from the Natural Sciences and Engineering Research Council of Canada (NSERC).

\bibliographystyle{apsrev4-2}
\bibliography{main}

@article{posti_2018,
	author = {{Posti, Lorenzo} and {Helmi, Amina}},
	title = {Mass and shape of the Milky Way’s dark matter halo with globular clusters from Gaia and Hubble},
	DOI= "10.1051/0004-6361/201833355",
	url= "https://doi.org/10.1051/0004-6361/201833355",
	journal = {A\&A},
	year = 2019,
	volume = 621,
	pages = "A56",
}

@article{Goldman_1989,
    author = {Goldman, Itzhak and Nussinov, Shmuel},
    title = {Weakly interacting massive particles and neutron stars},
    journal = {Physical Review D},
    year = {1989},
    month = {nov},
    volume = {40},
    number = {10},
    publisher={American Physical Society (APS)},
    pages = {3221--3230}
    
}

@article{Bell_2020,
   title={Improved treatment of dark matter capture in neutron stars},
   volume={2020},
   ISSN={1475-7516},
   url={http://dx.doi.org/10.1088/1475-7516/2020/09/028},
   DOI={10.1088/1475-7516/2020/09/028},
   number={09},
   journal={Journal of Cosmology and Astroparticle Physics},
   publisher={IOP Publishing},
   author={Bell, Nicole F. and Busoni, Giorgio and Robles, Sandra and Virgato, Michael},
   year={2020},
   month=sep, pages={028–028} 
}

@article{DeRocco_2022,
   title={Dark matter scattering in astrophysical media: collective effects},
   volume={2022},
   ISSN={1475-7516},
   url={http://dx.doi.org/10.1088/1475-7516/2022/05/015},
   DOI={10.1088/1475-7516/2022/05/015},
   number={05},
   journal={Journal of Cosmology and Astroparticle Physics},
   publisher={IOP Publishing},
   author={DeRocco, William and Galanis, Marios and Lasenby, Robert},
   year={2022},
   month=may, pages={015} 
}

@article{Bell_2021,
   title={Improved treatment of dark matter capture in neutron stars II:  leptonic  targets},
   volume={2021},
   ISSN={1475-7516},
   url={http://dx.doi.org/10.1088/1475-7516/2021/03/086},
   DOI={10.1088/1475-7516/2021/03/086},
   number={03},
   journal={Journal of Cosmology and Astroparticle Physics},
   publisher={IOP Publishing},
   author={Bell, Nicole F. and Busoni, Giorgio and Robles, Sandra and Virgato, Michael},
   year={2021},
   month=mar, pages={086} 
}

@article{Busoni_2022,
   title={Capture of Dark Matter in Neutron Stars},
   volume={77},
   ISSN={1934-8460},
   url={http://dx.doi.org/10.3103/S0027134922020205},
   DOI={10.3103/s0027134922020205},
   number={2},
   journal={Moscow University Physics Bulletin},
   publisher={Allerton Press},
   author={Busoni, Giorgio},
   year={2022},
   month=apr, pages={301–305} 
}

@misc{battaglieri2017cosmicvisionsnewideas,
      title={US Cosmic Visions: New Ideas in Dark Matter 2017: Community Report}, 
      author={Battaglieri et. al. , Marco},
      year={2017},
      eprint={1707.04591},
      archivePrefix={arXiv},
      primaryClass={hep-ph},
      url={https://arxiv.org/abs/1707.04591}, 
}

@article{Nelson_2019,
   title={Dark halos around neutron stars and gravitational waves},
   volume={2019},
   ISSN={1475-7516},
   url={http://dx.doi.org/10.1088/1475-7516/2019/07/012},
   DOI={10.1088/1475-7516/2019/07/012},
   number={07},
   journal={Journal of Cosmology and Astroparticle Physics},
   publisher={IOP Publishing},
   author={Nelson, Ann E. and Reddy, Sanjay and Zhou, Dake},
   year={2019},
   month=jul, pages={012–012} 
}

@article{de_Lavallaz_2010,
   title={Neutron stars as dark matter probes},
   volume={81},
   ISSN={1550-2368},
   url={http://dx.doi.org/10.1103/PhysRevD.81.123521},
   DOI={10.1103/physrevd.81.123521},
   number={12},
   journal={Physical Review D},
   publisher={American Physical Society (APS)},
   author={de Lavallaz, Arnaud and Fairbairn, Malcolm},
   year={2010},
   month=jun 
}

@article{Kouvaris_2010,
   title={Can neutron stars constrain dark matter?},
   volume={82},
   ISSN={1550-2368},
   url={http://dx.doi.org/10.1103/PhysRevD.82.063531},
   DOI={10.1103/physrevd.82.063531},
   number={6},
   journal={Physical Review D},
   publisher={American Physical Society (APS)},
   author={Kouvaris, Chris and Tinyakov, Peter},
   year={2010},
   month=sep 
}

@article{Bertone_2005,
   title={Time-dependent models for dark matter at the galactic center},
   volume={72},
   ISSN={1550-2368},
   url={http://dx.doi.org/10.1103/PhysRevD.72.103502},
   DOI={10.1103/physrevd.72.103502},
   number={10},
   journal={Physical Review D},
   publisher={American Physical Society (APS)},
   author={Bertone, Gianfranco and Merritt, David},
   year={2005},
   month=nov 
}

@article{McKeen_2018,
   title={Neutron Stars Exclude Light Dark Baryons},
   volume={121},
   ISSN={1079-7114},
   url={http://dx.doi.org/10.1103/PhysRevLett.121.061802},
   DOI={10.1103/physrevlett.121.061802},
   number={6},
   journal={Physical Review Letters},
   publisher={American Physical Society (APS)},
   author={McKeen, David and Nelson, Ann E. and Reddy, Sanjay and Zhou, Dake},
   year={2018},
   month=aug 
}

@article{Arvikar_2025,
   title={Exploring fermionic dark matter admixed neutron stars in the light of astrophysical observations},
   volume={112},
   ISSN={2470-0029},
   url={http://dx.doi.org/10.1103/kwwx-54wq},
   DOI={10.1103/kwwx-54wq},
   number={2},
   journal={Physical Review D},
   publisher={American Physical Society (APS)},
   author={Arvikar, Payaswinee and Gautam, Sakshi and Venneti, Anagh and Banik, Sarmistha},
   year={2025},
   month=jul 
}

@article{Merritt_2004,
  title = {Evolution of the Dark Matter Distribution at the Galactic Center},
  author = {Merritt, David},
  journal = {Phys. Rev. Lett.},
  volume = {92},
  issue = {20},
  pages = {201304},
  numpages = {4},
  year = {2004},
  month = {May},
  publisher = {American Physical Society},
  doi = {10.1103/PhysRevLett.92.201304},
  url = {https://link.aps.org/doi/10.1103/PhysRevLett.92.201304}
}

@article{Milosavljevic_2002,
   title={Galaxy cores as relics of black hole mergers},
   volume={331},
   ISSN={1365-2966},
   url={http://dx.doi.org/10.1046/j.1365-8711.2002.05436.x},
   DOI={10.1046/j.1365-8711.2002.05436.x},
   number={4},
   journal={Monthly Notices of the Royal Astronomical Society},
   publisher={Oxford University Press (OUP)},
   author={Milosavljevic, M. and Merritt, D. and Rest, A. and van den Bosch, F. C.},
   year={2002},
   month=apr, pages={L51–L55} 
}

@article{Ravindranath_2002,
   title={Nuclear Cusps and Cores in Early‐Type Galaxies as Relics of Binary Black Hole Mergers},
   volume={566},
   ISSN={1538-4357},
   url={http://dx.doi.org/10.1086/338228},
   DOI={10.1086/338228},
   number={2},
   journal={The Astrophysical Journal},
   publisher={American Astronomical Society},
   author={Ravindranath, Swara and Ho, Luis C. and Filippenko, Alexei V.},
   year={2002},
   month=feb, pages={801–808} 
}

@misc{khonji2024coreformationbinaryscouring,
      title={Core formation by binary scouring and gravitational wave recoil in massive elliptical galaxies}, 
      author={Nader Khonji and Alessia Gualandris and Justin I. Read and Walter Dehnen},
      year={2024},
      eprint={2408.12537},
      archivePrefix={arXiv},
      primaryClass={astro-ph.GA},
      url={https://arxiv.org/abs/2408.12537}, 
}

@article{10.1093/mnras/stab435,
    author = {Nasim, Imran Tariq and Gualandris, Alessia and Read, Justin I and Antonini, Fabio and Dehnen, Walter and Delorme, Maxime},
    title = {Formation of the largest galactic cores through binary scouring and gravitational wave recoil},
    journal = {Monthly Notices of the Royal Astronomical Society},
    volume = {502},
    number = {4},
    pages = {4794-4814},
    year = {2021},
    month = {02},
    issn = {0035-8711},
    doi = {10.1093/mnras/stab435},
    url = {https://doi.org/10.1093/mnras/stab435},
    eprint = {https://academic.oup.com/mnras/article-pdf/502/4/4794/36392316/stab435.pdf},
}

@article{Ilyin_2004,
   title={Dark matter in galaxies and the growth of giant black holes},
   volume={98},
   ISSN={1090-6509},
   url={http://dx.doi.org/10.1134/1.1648097},
   DOI={10.1134/1.1648097},
   number={1},
   journal={Journal of Experimental and Theoretical Physics},
   publisher={Pleiades Publishing Ltd},
   author={Ilyin, A. S. and Zybin, K. P. and Gurevich, A. V.},
   year={2004},
   month=jan, pages={1–13} 
}

@article{Gnedin_2004,
   title={Dark Matter Profile in the Galactic Center},
   volume={93},
   ISSN={1079-7114},
   url={http://dx.doi.org/10.1103/PhysRevLett.93.061302},
   DOI={10.1103/physrevlett.93.061302},
   number={6},
   journal={Physical Review Letters},
   publisher={American Physical Society (APS)},
   author={Gnedin, Oleg Y. and Primack, Joel R.},
   year={2004},
   month=aug 
}

@article{Wagner_2012,
   title={Torsion-balance tests of the weak equivalence principle},
   volume={29},
   ISSN={1361-6382},
   url={http://dx.doi.org/10.1088/0264-9381/29/18/184002},
   DOI={10.1088/0264-9381/29/18/184002},
   number={18},
   journal={Classical and Quantum Gravity},
   publisher={IOP Publishing},
   author={Wagner, T A and Schlamminger, S and Gundlach, J H and Adelberger, E G},
   year={2012},
   month=aug, pages={184002} 
}

@misc{cirelli2024darkmatter,
      title={Dark Matter}, 
      author={Marco Cirelli and Alessandro Strumia and Jure Zupan},
      year={2024},
      eprint={2406.01705},
      archivePrefix={arXiv},
      primaryClass={hep-ph},
      url={https://arxiv.org/abs/2406.01705}, 
}

@article{Ellis_2018,
   title={Dark matter effects on neutron star properties},
   volume={97},
   ISSN={2470-0029},
   url={http://dx.doi.org/10.1103/PhysRevD.97.123007},
   DOI={10.1103/physrevd.97.123007},
   number={12},
   journal={Physical Review D},
   publisher={American Physical Society (APS)},
   author={Ellis, John and Hütsi, Gert and Kannike, Kristjan and Marzola, Luca and Raidal, Martti and Vaskonen, Ville},
   year={2018},
   month=jun 
}

@inproceedings{Rafiei_Karkevandi_2023,
   title={Tidal deformability as a probe of dark matter in neutron stars},
   url={http://dx.doi.org/10.1142/9789811269776_0307},
   DOI={10.1142/9789811269776_0307},
   booktitle={The Sixteenth Marcel Grossmann Meeting},
   publisher={WORLD SCIENTIFIC},
   author={Rafiei Karkevandi, D. and Shakeri, S. and Sagun, V. and Ivanytskyi, O.},
   year={2023},
   month=jan, pages={3713–3731} 
}

@article{CIANCARELLA2021100796,
title = {Constraining mirror dark matter inside neutron stars},
journal = {Physics of the Dark Universe},
volume = {32},
pages = {100796},
year = {2021},
issn = {2212-6864},
doi = {https://doi.org/10.1016/j.dark.2021.100796},
url = {https://www.sciencedirect.com/science/article/pii/S2212686421000273},
author = {Raul Ciancarella and Francesco Pannarale and Andrea Addazi and Antonino Marcianò}
}

@article{gleason_2022,
  title = {Dynamical evolution of dark matter admixed neutron stars},
  author = {Gleason, Troy and Brown, Ben and Kain, Ben},
  journal = {Phys. Rev. D},
  volume = {105},
  issue = {2},
  pages = {023010},
  numpages = {14},
  year = {2022},
  month = {Jan},
  publisher = {American Physical Society},
  doi = {10.1103/PhysRevD.105.023010},
  url = {https://link.aps.org/doi/10.1103/PhysRevD.105.023010}
}

@article{flores_2024,
  title = {Gravitational wave asteroseismology of dark matter hadronic stars},
  author = {Flores, C\'esar V. and Lenzi, C. H. and Dutra, M. and Louren\ifmmode \mbox{\c{c}}\else \c{c}\fi{}o, O. and Arba\~nil, Jos\'e D. V.},
  journal = {Phys. Rev. D},
  volume = {109},
  issue = {8},
  pages = {083021},
  numpages = {12},
  year = {2024},
  month = {Apr},
  publisher = {American Physical Society},
  doi = {10.1103/PhysRevD.109.083021},
  url = {https://link.aps.org/doi/10.1103/PhysRevD.109.083021}
}

@article{das_2021,
  title = {Impacts of dark matter on the $f$-mode oscillation of hyperon star},
  author = {Das, H. C. and Kumar, Ankit and Biswal, S. K. and Patra, S. K.},
  journal = {Phys. Rev. D},
  volume = {104},
  issue = {12},
  pages = {123006},
  numpages = {14},
  year = {2021},
  month = {Dec},
  publisher = {American Physical Society},
  doi = {10.1103/PhysRevD.104.123006},
  url = {https://link.aps.org/doi/10.1103/PhysRevD.104.123006}
}

@article{shirke_2024,
  title = {Effects of dark matter on $f$-mode oscillations of neutron stars},
  author = {Shirke, Swarnim and Pradhan, Bikram Keshari and Chatterjee, Debarati and Sagunski, Laura and Schaffner-Bielich, J\"urgen},
  journal = {Phys. Rev. D},
  volume = {110},
  issue = {6},
  pages = {063025},
  numpages = {26},
  year = {2024},
  month = {Sep},
  publisher = {American Physical Society},
  doi = {10.1103/PhysRevD.110.063025},
  url = {https://link.aps.org/doi/10.1103/PhysRevD.110.063025}
}

@article{Garani_2021,
  title = {Observing the thermalization of dark matter in neutron stars},
  author = {Garani, Raghuveer and Gupta, Aritra and Raj, Nirmal},
  journal = {Phys. Rev. D},
  volume = {103},
  issue = {4},
  pages = {043019},
  numpages = {12},
  year = {2021},
  month = {Feb},
  publisher = {American Physical Society},
  doi = {10.1103/PhysRevD.103.043019},
  url = {https://link.aps.org/doi/10.1103/PhysRevD.103.043019}
}

@article{avila_2024,
    author = {Ávila, Afonso and Giangrandi, Edoardo and Sagun, Violetta and Ivanytskyi, Oleksii and Providência, Constança},
    title = {Rapid neutron star cooling triggered by dark matter},
    journal = {Monthly Notices of the Royal Astronomical Society},
    volume = {528},
    number = {4},
    pages = {6319-6328},
    year = {2024},
    month = {02},
    issn = {0035-8711},
    doi = {10.1093/mnras/stae337},
    url = {https://doi.org/10.1093/mnras/stae337},
    eprint = {https://academic.oup.com/mnras/article-pdf/528/4/6319/56719325/stae337.pdf},
}

@article{Bell_2024,
doi = {10.1088/1475-7516/2024/04/006},
url = {https://dx.doi.org/10.1088/1475-7516/2024/04/006},
year = {2024},
month = {apr},
publisher = {IOP Publishing},
volume = {2024},
number = {04},
pages = {006},
author = {Bell, Nicole F. and Busoni, Giorgio and Robles, Sandra and Virgato, Michael},
title = {Thermalization and annihilation of dark matter in neutron stars},
journal = {Journal of Cosmology and Astroparticle Physics}
}

@article{Kopp_2018,
   title={Cuckoo’s eggs in neutron stars: can LIGO hear chirps from the dark sector?},
   volume={2018},
   ISSN={1029-8479},
   url={http://dx.doi.org/10.1007/JHEP11(2018)096},
   DOI={10.1007/jhep11(2018)096},
   number={11},
   journal={Journal of High Energy Physics},
   publisher={Springer Science and Business Media LLC},
   author={Kopp, Joachim and Laha, Ranjan and Opferkuch, Toby and Shepherd, William},
   year={2018},
   month=nov 
}

@misc{owen_2025,
      title={Constraining dark-sector effects using gravitational waves from compact binary inspirals}, 
      author={Caroline B. Owen and Alexandria Tucker and Yonatan Kahn and Nicolás Yunes},
      year={2025},
      eprint={2503.04916},
      archivePrefix={arXiv},
      primaryClass={gr-qc},
      url={https://arxiv.org/abs/2503.04916}, 
}

@article{Alexander_2018,
   title={Hidden-sector modifications to gravitational waves from binary inspirals},
   volume={35},
   ISSN={1361-6382},
   url={http://dx.doi.org/10.1088/1361-6382/aaeb5c},
   DOI={10.1088/1361-6382/aaeb5c},
   number={23},
   journal={Classical and Quantum Gravity},
   publisher={IOP Publishing},
   author={Alexander, Stephon and McDonough, Evan and Sims, Robert and Yunes, Nicolás},
   year={2018},
   month=nov, pages={235012} 
}

@article{Cline_2018,
   title={Dark decay of the neutron},
   volume={2018},
   ISSN={1029-8479},
   url={http://dx.doi.org/10.1007/JHEP07(2018)081},
   DOI={10.1007/jhep07(2018)081},
   number={7},
   journal={Journal of High Energy Physics},
   publisher={Springer Science and Business Media LLC},
   author={Cline, James M. and Cornell, Jonathan M.},
   year={2018},
   month=jul 
}

@misc{basterogil_2024,
      title={The neutron decay anomaly, neutron stars and dark matter}, 
      author={Mar Bastero-Gil and Teresa Huertas-Roldan and Daniel Santos},
      year={2024},
      eprint={2403.08666},
      archivePrefix={arXiv},
      primaryClass={astro-ph.CO},
      url={https://arxiv.org/abs/2403.08666}, 
}

@article{Grinstein_2019,
   title={Neutron Star Stability in Light of the Neutron Decay Anomaly},
   volume={123},
   ISSN={1079-7114},
   url={http://dx.doi.org/10.1103/PhysRevLett.123.091601},
   DOI={10.1103/physrevlett.123.091601},
   number={9},
   journal={Physical Review Letters},
   publisher={American Physical Society (APS)},
   author={Grinstein, Benjamín and Kouvaris, Chris and Nielsen, Niklas Grønlund},
   year={2019},
   month=aug 
}

@article{Bell:2018pkk,
    author = "Bell, Nicole F. and Busoni, Giorgio and Robles, Sandra",
    title = "{Heating up Neutron Stars with Inelastic Dark Matter}",
    eprint = "1807.02840",
    archivePrefix = "arXiv",
    primaryClass = "hep-ph",
    doi = "10.1088/1475-7516/2018/09/018",
    journal = "JCAP",
    volume = "09",
    pages = "018",
    year = "2018"
}

@article{Sadeghian_2013,
   title={Dark-matter distributions around massive black holes: A general relativistic analysis},
   volume={88},
   ISSN={1550-2368},
   url={http://dx.doi.org/10.1103/PhysRevD.88.063522},
   DOI={10.1103/physrevd.88.063522},
   number={6},
   journal={Physical Review D},
   publisher={American Physical Society (APS)},
   author={Sadeghian, Laleh and Ferrer, Francesc and Will, Clifford M.},
   year={2013},
   month=sep 
}

@article{Bertone_2025,
   title={Toward a realistic description of dark matter overdensities around black holes},
   volume={112},
   ISSN={2470-0029},
   url={http://dx.doi.org/10.1103/5nnf-8fz9},
   DOI={10.1103/5nnf-8fz9},
   number={4},
   journal={Physical Review D},
   publisher={American Physical Society (APS)},
   author={Bertone, Gianfranco and Wierda, A. Renske A. C. and Gaggero, Daniele and Kavanagh, Bradley J. and Volonteri, Marta and Yoshida, Naoki},
   year={2025},
   month=aug 
}

@article{Kaplinghat_2016,
   title={Dark Matter Halos as Particle Colliders: Unified Solution to Small-Scale Structure Puzzles from Dwarfs to Clusters},
   volume={116},
   ISSN={1079-7114},
   url={http://dx.doi.org/10.1103/PhysRevLett.116.041302},
   DOI={10.1103/physrevlett.116.041302},
   number={4},
   journal={Physical Review Letters},
   publisher={American Physical Society (APS)},
   author={Kaplinghat, Manoj and Tulin, Sean and Yu, Hai-Bo},
   year={2016},
   month=jan
}

@article{Tulin_2018,
   title={Dark matter self-interactions and small scale structure},
   volume={730},
   ISSN={0370-1573},
   url={http://dx.doi.org/10.1016/j.physrep.2017.11.004},
   DOI={10.1016/j.physrep.2017.11.004},
   journal={Physics Reports},
   publisher={Elsevier BV},
   author={Tulin, Sean and Yu, Hai-Bo},
   year={2018},
   month=feb, pages={1–57} }

@article{Braaten_2018,
   title={Production of dark-matter bound states in the early universe by three-body recombination},
   volume={2018},
   ISSN={1029-8479},
   url={http://dx.doi.org/10.1007/JHEP11(2018)084},
   DOI={10.1007/jhep11(2018)084},
   number={11},
   journal={Journal of High Energy Physics},
   publisher={Springer Science and Business Media LLC},
   author={Braaten, Eric and Kang, Daekyoung and Laha, Ranjan},
   year={2018},
   month=nov }

@misc{roy_2024,
      title={Aggressively-Dissipative Dark Dwarfs: The Effects of Atomic Dark Matter on the Inner Densities of Isolated Dwarf Galaxies}, 
      author={Sandip Roy and Xuejian Shen and Jared Barron and Mariangela Lisanti and David Curtin and Norman Murray and Philip F. Hopkins},
      year={2024},
      eprint={2408.15317},
      archivePrefix={arXiv},
      primaryClass={astro-ph.GA},
      url={https://arxiv.org/abs/2408.15317}, 
}

@article{FAN_2013,
title = {Double-Disk Dark Matter},
journal = {Physics of the Dark Universe},
volume = {2},
number = {3},
pages = {139-156},
year = {2013},
issn = {2212-6864},
doi = {https://doi.org/10.1016/j.dark.2013.07.001},
url = {https://www.sciencedirect.com/science/article/pii/S2212686413000289},
author = {JiJi Fan and Andrey Katz and Lisa Randall and Matthew Reece}
}

@article{Alonso_alvarez_2024,
   title={Baryogenesis through asymmetric reheating in the mirror twin Higgs},
   volume={2024},
   ISSN={1029-8479},
   url={http://dx.doi.org/10.1007/JHEP05(2024)069},
   DOI={10.1007/jhep05(2024)069},
   number={5},
   journal={Journal of High Energy Physics},
   publisher={Springer Science and Business Media LLC},
   author={Alonso-Álvarez, Gonzalo and Curtin, David and Rasovic, Andrija and Yuan, Zhihan},
   year={2024},
   month=may }

@article{Beauchesne_2021,
   title={Cosmology of the Twin Higgs without explicit ℤ2 breaking},
   volume={2021},
   ISSN={1029-8479},
   url={http://dx.doi.org/10.1007/JHEP12(2021)160},
   DOI={10.1007/jhep12(2021)160},
   number={12},
   journal={Journal of High Energy Physics},
   publisher={Springer Science and Business Media LLC},
   author={Beauchesne, Hugues and Kats, Yevgeny},
   year={2021},
   month=dec }

@article{McCullough_2013,
doi = {10.1088/1475-7516/2013/10/058},
url = {https://doi.org/10.1088/1475-7516/2013/10/058},
year = {2013},
month = {oct},
publisher = {},
volume = {2013},
number = {10},
pages = {058},
author = {Matthew McCullough and Lisa Randall},
title = {Exothermic double-disk dark matter},
journal = {Journal of Cosmology and Astroparticle Physics}
}

@article{Gurian_2022,
doi = {10.3847/1538-4357/ac75e4},
url = {https://doi.org/10.3847/1538-4357/ac75e4},
year = {2022},
month = {aug},
publisher = {The American Astronomical Society},
volume = {934},
number = {2},
pages = {121},
author = {Gurian, James and Jeong, Donghui and Ryan, Michael and Shandera, Sarah},
title = {Molecular Chemistry for Dark Matter. II. Recombination, Molecule Formation, and Halo Mass Function in Atomic Dark Matter},
journal = {The Astrophysical Journal}
}

@article{Agrawal_2017,
doi = {10.1088/1475-7516/2017/05/022},
url = {https://doi.org/10.1088/1475-7516/2017/05/022},
year = {2017},
month = {may},
publisher = {},
volume = {2017},
number = {05},
pages = {022},
author = {Agrawal, Prateek and Cyr-Racine, Francis-Yan and Randall, Lisa and Scholtz, Jakub},
title = {Make dark matter charged again},
journal = {Journal of Cosmology and Astroparticle Physics}
}

@article{Abbott_2017,
   title={GW170817: Observation of Gravitational Waves from a Binary Neutron Star Inspiral},
   volume={119},
   ISSN={1079-7114},
   url={http://dx.doi.org/10.1103/PhysRevLett.119.161101},
   DOI={10.1103/physrevlett.119.161101},
   number={16},
   journal={Physical Review Letters},
   publisher={American Physical Society (APS)},
   author={Abbott et. al, B. P.},
   year={2017},
   month=oct,
}

@misc{scarcella2025hintsdarkmatterspikes,
      title={Hints of Dark Matter Spikes in Low-mass X-ray Binaries: a critical assessment}, 
      author={Francesca Scarcella and Bradley J. Kavanagh},
      year={2025},
      eprint={2510.11635},
      archivePrefix={arXiv},
      primaryClass={hep-ph},
      url={https://arxiv.org/abs/2510.11635}, 
}

@article{Berg__2022,
   title={MICROSCOPE’s constraint on a short-range fifth force},
   volume={39},
   ISSN={1361-6382},
   url={http://dx.doi.org/10.1088/1361-6382/abe142},
   DOI={10.1088/1361-6382/abe142},
   number={20},
   journal={Classical and Quantum Gravity},
   publisher={IOP Publishing},
   author={Bergé, Joel and Pernot-Borràs, Martin and Uzan, Jean-Philippe and Brax, Philippe and Chhun, Ratana and Métris, Gilles and Rodrigues, Manuel and Touboul, Pierre},
   year={2022},
   month=sep, pages={204010} 
}

@article{Kapner_2007,
   title={Tests of the Gravitational Inverse-Square Law below the Dark-Energy Length Scale},
   volume={98},
   ISSN={1079-7114},
   url={http://dx.doi.org/10.1103/PhysRevLett.98.021101},
   DOI={10.1103/physrevlett.98.021101},
   number={2},
   journal={Physical Review Letters},
   publisher={American Physical Society (APS)},
   author={Kapner, D. J. and Cook, T. S. and Adelberger, E. G. and Gundlach, J. H. and Heckel, B. R. and Hoyle, C. D. and Swanson, H. E.},
   year={2007},
   month=jan 
}

@mastersthesis{jockel_2023,
author = {Jockel, Cédric},
year = {2023},
school = {Goethe Universit\"{a}t},
month = {08},
pages = {},
title = {Scalar- and Vector Dark Matter Admixed Neutron Stars},
doi = {10.48550/arXiv.2308.12174}
}

@article{rutherford_2023,
  title = {Constraining bosonic asymmetric dark matter with neutron star mass-radius measurements},
  author = {Rutherford, Nathan and Raaijmakers, Geert and Prescod-Weinstein, Chanda and Watts, Anna},
  journal = {Phys. Rev. D},
  volume = {107},
  issue = {10},
  pages = {103051},
  numpages = {19},
  year = {2023},
  month = {May},
  publisher = {American Physical Society},
  doi = {10.1103/PhysRevD.107.103051},
  url = {https://link.aps.org/doi/10.1103/PhysRevD.107.103051}
}

@article{colpi_1986,
  title = {Boson Stars: Gravitational Equilibria of Self-Interacting Scalar Fields},
  author = {Colpi, Monica and Shapiro, Stuart L. and Wasserman, Ira},
  journal = {Phys. Rev. Lett.},
  volume = {57},
  issue = {20},
  pages = {2485--2488},
  numpages = {0},
  year = {1986},
  month = {Nov},
  publisher = {American Physical Society},
  doi = {10.1103/PhysRevLett.57.2485},
  url = {https://link.aps.org/doi/10.1103/PhysRevLett.57.2485}
}

@misc{cong2024spindependentexoticinteractions,
      title={Spin-dependent exotic interactions}, 
      author={Lei Cong and Wei Ji and Pavel Fadeev and Filip Ficek and Min Jiang and Victor V. Flambaum and Haosen Guan and Derek F. Jackson Kimball and Mikhail G. Kozlov and Yevgeny V. Stadnik and Dmitry Budker},
      year={2024},
      eprint={2408.15691},
      archivePrefix={arXiv},
      primaryClass={hep-ph},
      doi={https://doi.org/10.1103/RevModPhys.97.025005},
      url={https://arxiv.org/abs/2408.15691}, 
}

@article{PhysRevD.105.023001,
  title = {Bosonic dark matter in neutron stars and its effect on gravitational wave signal},
  author = {Rafiei Karkevandi, Davood and Shakeri, Soroush and Sagun, Violetta and Ivanytskyi, Oleksii},
  journal = {Phys. Rev. D},
  volume = {105},
  issue = {2},
  pages = {023001},
  numpages = {18},
  year = {2022},
  month = {Jan},
  publisher = {American Physical Society},
  doi = {10.1103/PhysRevD.105.023001},
  url = {https://link.aps.org/doi/10.1103/PhysRevD.105.023001}
}

@article{Dror:2017ehi,
    author = "Dror, Jeff A. and Lasenby, Robert and Pospelov, Maxim",
    title = "{New constraints on light vectors coupled to anomalous currents}",
    eprint = "1705.06726",
    archivePrefix = "arXiv",
    primaryClass = "hep-ph",
    doi = "10.1103/PhysRevLett.119.141803",
    journal = "Phys. Rev. Lett.",
    volume = "119",
    number = "14",
    pages = "141803",
    year = "2017"
}

@article{Dror:2017nsg,
    author = "Dror, Jeff A. and Lasenby, Robert and Pospelov, Maxim",
    title = "{Dark forces coupled to nonconserved currents}",
    eprint = "1707.01503",
    archivePrefix = "arXiv",
    primaryClass = "hep-ph",
    doi = "10.1103/PhysRevD.96.075036",
    journal = "Phys. Rev. D",
    volume = "96",
    number = "7",
    pages = "075036",
    year = "2017"
}

@techreport{Drlica-Wagner_2022,
    author = "Drlica-Wagner, Alex and others",
    title = "{Report of the Topical Group on Cosmic Probes of Dark Matter for Snowmass 2021}",
    eprint = "2209.08215",
    archivePrefix = "arXiv",
    primaryClass = "hep-ph",
    reportNumber = "FERMILAB-FN-1211-PPD",
    doi = "10.2172/1898829",
    institution = "SLAC National Accelerator Laboratory",
    month = "sep",
    year = "2022"
}

@inproceedings{Akerib:2022ort,
    author = "Akerib, D. S. and others",
    title = "{Snowmass2021 Cosmic Frontier Dark Matter Direct Detection to the Neutrino Fog}",
    booktitle = "{Snowmass 2021}",
    eprint = "2203.08084",
    archivePrefix = "arXiv",
    primaryClass = "hep-ex",
    reportNumber = "FERMILAB-CONF-22-180-V",
    month = "3",
    year = "2022"
}

@inproceedings{Essig:2022dfa,
    author = "Essig, Rouven and others",
    title = "{Snowmass2021 Cosmic Frontier: The landscape of low-threshold dark matter direct detection in the next decade}",
    booktitle = "{Snowmass 2021}",
    eprint = "2203.08297",
    archivePrefix = "arXiv",
    primaryClass = "hep-ph",
    reportNumber = "FERMILAB-CONF-22-181-PPD",
    month = "3",
    year = "2022"
}

@article{Kaplan:2009de,
    author = "Kaplan, David E. and Krnjaic, Gordan Z. and Rehermann, Keith R. and Wells, Christopher M.",
    title = "{Atomic Dark Matter}",
    eprint = "0909.0753",
    archivePrefix = "arXiv",
    primaryClass = "hep-ph",
    doi = "10.1088/1475-7516/2010/05/021",
    journal = "JCAP",
    volume = "05",
    pages = "021",
    year = "2010"
}

@misc{graham2025cosmologicallimitsstrongdark,
      title={Cosmological Limits on Strong Dark Forces}, 
      author={Peter W. Graham and Harikrishnan Ramani and Olivier Simon and Erwin H. Tanin},
      year={2025},
      eprint={2511.09614},
      archivePrefix={arXiv},
      primaryClass={hep-ph},
      url={https://arxiv.org/abs/2511.09614}, 
}

@article{Leane_2021,
   title={Celestial-body focused dark matter annihilation throughout the Galaxy},
   volume={103},
   ISSN={2470-0029},
   url={http://dx.doi.org/10.1103/PhysRevD.103.075030},
   DOI={10.1103/physrevd.103.075030},
   number={7},
   journal={Physical Review D},
   publisher={American Physical Society (APS)},
   author={Leane, Rebecca K. and Linden, Tim and Mukhopadhyay, Payel and Toro, Natalia},
   year={2021},
   month=apr 
}

@article{Acevedo_2020,
   title={Warming nuclear pasta with dark matter: kinetic and annihilation heating of neutron star crusts},
   volume={2020},
   ISSN={1475-7516},
   url={http://dx.doi.org/10.1088/1475-7516/2020/03/038},
   DOI={10.1088/1475-7516/2020/03/038},
   number={03},
   journal={Journal of Cosmology and Astroparticle Physics},
   publisher={IOP Publishing},
   author={Acevedo, Javier F. and Bramante, Joseph and Leane, Rebecca K. and Raj, Nirmal},
   year={2020},
   month=mar, pages={038–038} 
}

@article{Yunes:2009ke,
    author = "Yunes, Nicolas and Pretorius, Frans",
    title = "{Fundamental Theoretical Bias in Gravitational Wave Astrophysics and the Parameterized Post-Einsteinian Framework}",
    eprint = "0909.3328",
    archivePrefix = "arXiv",
    primaryClass = "gr-qc",
    doi = "10.1103/PhysRevD.80.122003",
    journal = "Phys. Rev. D",
    volume = "80",
    pages = "122003",
    year = "2009"
}

@article{Cornish:2011ys,
    author = "Cornish, Neil and Sampson, Laura and Yunes, Nicolas and Pretorius, Frans",
    title = "{Gravitational Wave Tests of General Relativity with the Parameterized Post-Einsteinian Framework}",
    eprint = "1105.2088",
    archivePrefix = "arXiv",
    primaryClass = "gr-qc",
    doi = "10.1103/PhysRevD.84.062003",
    journal = "Phys. Rev. D",
    volume = "84",
    pages = "062003",
    year = "2011"
}

@article{Yunes:2016jcc,
    author = "Yunes, Nicolas and Yagi, Kent and Pretorius, Frans",
    title = "{Theoretical Physics Implications of the Binary Black-Hole Mergers GW150914 and GW151226}",
    eprint = "1603.08955",
    archivePrefix = "arXiv",
    primaryClass = "gr-qc",
    doi = "10.1103/PhysRevD.94.084002",
    journal = "Phys. Rev. D",
    volume = "94",
    number = "8",
    pages = "084002",
    year = "2016"
}

@misc{giffin2025structureformationdarkmagnetohydrodynamics,
      title={Structure Formation with Dark Magnetohydrodynamics}, 
      author={Pierce Giffin and Andrew Liu and Jeremias Boucsein and Akaxia Cruz and Anirudh Prabhu and Stefano Profumo and M. Grant Roberts},
      year={2025},
      eprint={2511.15810},
      archivePrefix={arXiv},
      primaryClass={hep-ph},
      url={https://arxiv.org/abs/2511.15810}, 
}

@article{Cruz_2023,
   title={Astrophysical plasma instabilities induced by long-range interacting dark matter},
   volume={2023},
   ISSN={1475-7516},
   url={http://dx.doi.org/10.1088/1475-7516/2023/04/028},
   DOI={10.1088/1475-7516/2023/04/028},
   number={04},
   journal={Journal of Cosmology and Astroparticle Physics},
   publisher={IOP Publishing},
   author={Cruz, Akaxia and McQuinn, Matthew},
   year={2023},
   month=apr, pages={028} 
}

@article{DeRocco_2025,
   title={Dark plasmas in the nonlinear regime: Constraints from particle-in-cell simulations},
   volume={111},
   ISSN={2470-0029},
   url={http://dx.doi.org/10.1103/PhysRevD.111.095031},
   DOI={10.1103/physrevd.111.095031},
   number={9},
   journal={Physical Review D},
   publisher={American Physical Society (APS)},
   author={DeRocco, William and Giffin, Pierce},
   year={2025},
   month=may 
}
\end{document}